\documentclass[prd,amsmath,amssymb,superscriptaddress,nofootinbib,twocolumn]{revtex4}
\usepackage{amsfonts}
\usepackage{multirow}
\usepackage{makecell}
\usepackage{mathrsfs}
\usepackage{graphicx}
\usepackage{amsmath}
\usepackage{amssymb}
\usepackage{bm}
\usepackage{bbm}
\usepackage{color}
\usepackage{ulem}
\usepackage{array}
\usepackage{diagbox}
\usepackage{float}

\newcommand{\nn}{\nonumber}

\newcommand{\beq}{\begin{equation}}
\newcommand{\eeq}{\end{equation}}
\newcommand{\bqa}{\begin{eqnarray}}
\newcommand{\eqa}{\end{eqnarray}}

\newcommand{\bseq}{\begin{subequations}}
\newcommand{\eseq}{\end{subequations}}

\makeatletter

\begin{document}

\title{Next-to-leading-order electroweak correction to $H\to Z^0\gamma$
}
\author{Wen-Long Sang~\footnote{wlsang@swu.edu.cn}}
\affiliation{School of Physical Science and Technology, Southwest University, Chongqing 400700, China\vspace{0.2cm}}
\author{Feng Feng~\footnote{f.feng@outlook.com}}
\affiliation{China University of Mining and Technology, Beijing 100083, China\vspace{0.2cm}}
\affiliation{Institute of High Energy Physics,
Chinese Academy of Sciences, Beijing 100049, China\vspace{0.2cm}}
\author{Yu Jia~\footnote{jiay@ihep.ac.cn}}
\affiliation{Institute of High Energy Physics, Chinese Academy of Sciences, Beijing 100049, China\vspace{0.2cm}}
\affiliation{School of Physical Sciences, University of Chinese Academy of Sciences, Beijing 100049, China\vspace{0.2cm}}

\date{\today}

\begin{abstract}
Inspired by the recent observation of the Higgs boson radiative decay into $Z^0$ by {\tt ATLAS} and {\tt CMS} Collaborations,
we investigate the next-to-leading-order (NLO) electroweak correction to this rare decay process in Standard Model (SM).
Implementing the on-shell renormalization scheme, we find that the magnitude of the
NLO electroweak correction may reach $7\%$ of the leading order (LO) prediction,
much more significant than that of the NLO QCD correction, which is merely about $0.3\%$.
After incorporating the ${\cal O}(\alpha)$ correction, the predicted partial width from various $\alpha$ schemes
tend to converge to each other. Including both NLO electroweak and QCD corrections,
the SM prediction for the branching fraction shifts from the LO value of $(1.40-1.71)\times 10^{-3}$ to
$(1.55\pm 0.06)\times 10^{-3}$, considerably lower than the measured value ${\cal B}_{\rm exp}[H\to Z^0\gamma]=(3.4\pm 1.1)\times 10^{-3}$.
Resolving this discrepancy clearly calls for further theoretical investigations, and, more importantly,
experimental efforts from {\tt HL-LHC} and the prospective Higgs factories such as {\tt CEPC} and {\tt FCC-ee}.
\end{abstract}

\maketitle

\noindent{\color{blue}\it Introduction.}  The ground-breaking discovery of a 125 GeV boson at {\tt LHC} in 2012
heralds a new era of particle physics~\cite{ATLAS:2012yve,CMS:2012qbp}.
After decade long experimental endeavours, the couplings between this new particle and $W$, $Z$, heavy fermions
have been confirmed to be consistent with the Standard Model (SM) predictions,
provided that this new boson is identified with the Higgs boson~\cite{ATLAS:2016neq,ATLAS:2022vkf,CMS:2022dwd}.
Nevertheless, it is still important to test the SM predictions in some rare decay channels of Higgs boson,
exemplified by  $H\to Z^0\gamma$. As a loop-induced process, this decay channel serves a fruitful platform
to look for the footprints of new particles predicted in many beyond Standard Model (BSM) scenarios.

Recently, the {\tt ATLAS} and {\tt CMS} Collaborations jointly reported the evidence of observing the rare decay $H\to Z^0\gamma$~\cite{ATLAS:2023yqk,ATLAS:2020qcv,CMS:2022ahq},
and found $\mu_{Z\gamma}=2.2\pm 0.7$, with $\mu\equiv (\sigma {\cal B})_{\rm meas}/(\sigma {\cal B})_{\rm th}$ being a parameter to gauge how well
the measurement conforms to the SM prediction. The measured branching fraction is ${\cal B}(H\to Z\gamma) =(3.4\pm 1.1)\times 10^{-3}$,
more than twice larger than the SM prediction $(1.5\pm 0.1)\times 10^{-3}$~\cite{Djouadi:1997yw,LHCHiggsCrossSectionWorkingGroup:2016ypw}.
This should be contrasted with a similar channel $H\to \gamma\gamma$ with $\mu_{\gamma\gamma}=1.04^{+0.10}_{-0.09}$~\cite{ATLAS:2022tnm},
which exhibits a satisfactory agreement between
the measurement and the SM prediction.
Recently there has appeared a flurry of theoretical work
attempting to alleviate the discrepancy between  the measurements and SM predictions for $H\to Z^0\gamma$,
by invoking various BSM models~\cite{Herrero:2020dtv,Hong:2023mwr,Panghal:2023iqd,Buccioni:2023qnt,Barducci:2023zml,Boto:2023bpg,Cheung:2024kml,He:2024sma}.

\begin{figure}[hbtp]
\centering
\includegraphics[width=0.5\textwidth]{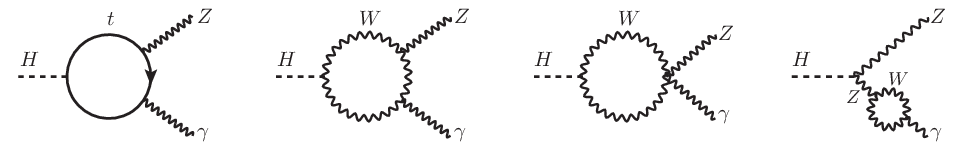}
\caption{Some representative LO diagrams for $H\to Z^0\gamma$.}
\label{Feynman:Diagram:LO}
\end{figure}

Needless to say, in order to ultimately resolve the discrepancy related to $H\to Z^0\gamma$,
it is compulsive to first have a as precise as possible prediction in SM.
At leading order (LO), the process
$H\to Z^0\gamma$ is mediated via a heavy quark or a $W$ boson loop, as depicted in Fig.~\ref{Feynman:Diagram:LO}.
The LO prediction has been available four decades ago~\cite{Cahn:1978nz,Bergstrom:1985hp}.
It turns out that the $W$-loop induced contribution dominates the $t$-loop induced one,
while there exists a destructive interference between these two channels.
The NLO QCD correction in numerical form has been known in early 90s~\cite{Spira:1991tj},
and its analytical form has also been available about one decade ago~\cite{Gehrmann:2015dua,Bonciani:2015eua}.
The QCD correction appears to be rather insignificant. This may be partly understood by the fact that the
gluons can be attached only to the top quark loop, which by itself only yields
a small portion of the contribution with respect to the $W$ loop.

Therefore, the missing NLO electroweak correction is envisaged to constitute the major theoretical uncertainty
for this decay channel, which has been estimated to reach $5\%$ level of the LO contribution~\cite{LHCHiggsCrossSectionWorkingGroup:2016ypw}.
It is the goal of this work to conduct a detailed investigation on the
NLO electroweak correction for $H\to Z^0\gamma$.

\vspace{0.2cm}

\noindent{\color{blue}\it Partial width from form factors.}
Let us express the amplitude for $H\to Z^0\gamma$ as $\mathcal{A}=T_{\mu\nu}\varepsilon^{\mu*}_Z(p_1) \varepsilon^{\nu*}_\gamma(p_2)$.
By Lorentz invariance, $T_{\mu\nu}$ can be decomposed into the following most general tensor structures:
\bqa
 T^{\mu\nu} &=& p_1^\mu p_1^\nu T_1+p_2^\mu p_2^\nu T_2+p_1^\mu p_2^\nu T_3+p_2^\mu p_1^\nu T_4
\nn\\
&+& p_1\cdot p_2 g^{\mu\nu} T_5 +
\epsilon^{\mu\nu\alpha\beta}p_{1\alpha}p_{2\beta}T_6,
\label{eq:lorentz:structure}
\eqa
where $p_1$ and $p_2$ signify the momenta of the outgoing $Z$ boson and photon, respectively, and $T_i$ ($i=1,\cdots 6$) signify six scalar form factors.
The transversity condition $p_1\cdot \epsilon_Z=p_2\cdot \epsilon_\gamma=0$ implies that the terms entailing
$T_2$ and $T_3$ do not contribute. Ward identity $p_{2\nu}T^{\mu\nu}=0$ implies that $T_1=0$ and $T_4=-T_5$.
The $CP$-violating $T_6$ term first arises at two-loop order in SM, but its interference with the LO amplitude yields a vanishing contribution.
Hence we can neglect the $T_6$ term at the outset as far as the NLO electroweak correction is concerned.
Consequently, only one independent form factor, $T_5=-T_4$, finally survives.
In practice, it is convenient to introduce a new dimensionless form factor $\widetilde{T}_5$:
\beq
T_5=\frac{M_H^2}{M_W s_W(M_H^2-M_Z^2)}\widetilde{T}_5,
\eeq
where $c_W\equiv \tfrac{M_W}{M_Z}$, and $s_W=\sqrt{1-c_W^2}$.
The partial width then becomes
\beq
\Gamma(H\to Z^0\gamma)=\frac{1}{2M_H}\frac{1}{8\pi}\frac{2|{\bf p}_1|}{M_H}\frac{M_H^4}{2 M_W^2 s_W^2}|\widetilde{T}_5|^2,
\label{Partial:width:from:T5:form:factor}
\eeq
where the three-momentum of the $Z^0$ boson in the Higgs boson rest frame is $|{\bf p}_1|={(M_H^2-M_Z^2)}/{2M_H}$.
Our task is to deduce $\widetilde{T}_5$ through NLO in $\alpha$.

\vspace{0.2cm}

\begin{figure}[hbtp]
	\centering
	\includegraphics[width=0.5\textwidth]{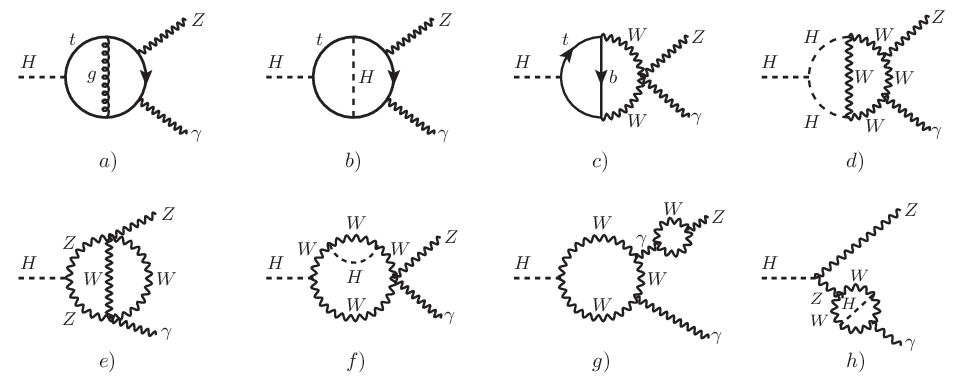}
	\caption{Some representative two-loop diagrams for $H\to Z^0\gamma$.
     $a)$ represents a sample diagram for NLO QCD correction, while $b)$-$h)$ represent sample
     diagrams for NLO electroweak correction.}
	\label{Feynman:Diagram:NLO}
\end{figure}

\noindent{\color{blue}\it  Outline of calculation.}
Throughout this work we adopt the Feynman gauge and employ the dimensional regularization to regularize UV divergences.
We utilize the package {\tt FeynArts}~\cite{Hahn:2000kx} to generate Feynman diagrams and the corresponding amplitudes for
$H\to Z^0\gamma$ through NLO in $\alpha$.
This process entails approximately 50 LO (one-loop) diagrams, and more than $10^4$ NLO (two-loop) diagrams~(including the NLO QCD diagrams).
Some representative diagrams relevant to the NLO QCD and electroweak corrections are depicted in Fig.~\ref{Feynman:Diagram:NLO}.
The form factors $T_4$ and $T_5$ are extracted from the amplitude with the aid of the covariant projectors~\cite{Bonciani:2015eua}.
We employ the packages {\tt FeynCalc}~\cite{Mertig:1990an} and {\tt FormLink}~\cite{Feng:2012tk} to perform Lorentz contraction and Dirac trace.
We use the packages {\tt Apart}~\cite{Feng:2012iq} and {\tt FIRE}~\cite{Smirnov:2014hma} for integration-by-part reduction.
We end up with 8 one-loop master integrals (MIs) and over 700 two-loop MIs.
The package {\tt AMFlow}~\cite{Liu:2017jxz,Liu:2022mfb,Liu:2022chg} is invoked to evaluate all these MIs with high numerical accuracy.

The on-shell renormalization scheme\footnote{
We stress that 
the widths of $W/Z$ bosons have been neglected, because these virtual particles can not become on shell simultaneously.
In principle, one may adopt the complex mass scheme~\cite{Denner:1999gp,Denner:2006ic} 
to incorporate the width effect in a gauge invariant way. The difference between the complex mass scheme and the conventional on-shell renormalization scheme is anticipated to be immaterial at the level of accuracy concerned in this work.} has been widely used in the field of electroweak radiative correction to tame the UV divergence~\cite{Ross:1973fp,Hollik:1988ii}.
The renormalized parameters are chosen to be those measured very precisely,
such as Higgs boson mass, $W/Z$ masses, top quark mass, and QED coupling at Thomson limit.
Having implemented the mass and charge renormalization ($e_0=Z_e e$), we find it convenient to stay with the bare field
in the practical calculation.
Following the LSZ reduction formula, we multiply the amputated amplitude of $H \to Z^0\gamma$ by the factor
$\sqrt{ Z_H Z_{ZZ} Z_{AA}}$, where the $Z_H$, $Z_{ZZ}$ and $Z_{AA}$ denote the field renormalization
constants associated with $H$, $Z^0$ and $\gamma$, respectively.
These field renormalization constants can be inferred from the unrenormalized propagators to one loop accuracy,
by identifying the residues in the on-shell limit.
It is important to explicitly include those diagrams where $Z^0$  converts via loop into a outgoing photon in the external leg,
or vice versa,  as depicted in Fig.~\ref{Feynman:Diagram:LO}$d)$ and Fig.~\ref{Feynman:Diagram:NLO}$g)$, $h)$~\footnote{Alternatively,
one may choose to work with the renormalized gauge fields and introduce new counterterms in the renormalized SM Lagrangian.
The renormalization constants linking the bare and renormalized $Z^0$ and $\gamma$ fields have to be a $2\times 2$
matrix with non-vanishing off-diagonal elements~\cite{Denner:1991kt}:
\bqa
\begin{pmatrix}
Z_0\\
A_0
\end{pmatrix}
=
\begin{pmatrix}
1+\frac{1}{2}\delta Z_{ZZ} &\frac{1}{2}\delta Z_{ZA}\\
\frac{1}{2}\delta Z_{AZ} & 1+\frac{1}{2}\delta Z_{AA}
\end{pmatrix}
\begin{pmatrix}
Z\\
A
\end{pmatrix}
.\nn
\eqa
By the on-shell renormalization condition, in this case one no longer needs compute those topologically unamputated diagrams
exemplified by Fig.~\ref{Feynman:Diagram:LO}$d)$ and Fig.~\ref{Feynman:Diagram:NLO}$g)$, $h)$.
However, as an extra price to pay, one has to consider the $HZ\gamma$ vertex induced by the counterterm lagrangian,
with a Feynman rule $ie g_{\mu\nu} M_W{1\over s_W c_W^2}{1\over 2}\delta Z_{ZA}$~\cite{Denner:1991kt}.
It is envisaged that the renormalized perturbation theory must yield the identical results as the bare field approach as adopted in this work.}.

After carrying out the mass and charge renormalization, we finally end up with the UV-finite results for $T_4$ and $T_5$.
We have confirmed the Ward identity requirement $T_4=-T_5$ holds at ${\cal O}(\alpha)$ level, which serves a nontrivial crosscheck.

\vspace{0.2cm}

\noindent{\color{blue}\it Three different recipes about charge renormalization.}
Implementing the charge renormalization in electroweak theory is not unique,
which leads to several popular variants in the on-shell renormalization scheme.
In the so-called $\alpha(0)$ scheme, where $\alpha$ is taking with the Thomson-limit value.
To one-loop order, $\delta Z_e$ is expressed as
\beq\label{eq:ze:orig}
\delta Z_e|_{\alpha(0)}= {1\over 2} \Pi^{AA}(0)-{s_W\over c_W}
{\Sigma^{A Z}_T(0) \over M_Z^2},
\eeq
where ${\Pi(s)}\equiv \Sigma_T^{AA}(s)/s$.
The photon vacuum polarization in small momentum transfer is sensitive to the low-energy hadronic contribution,
thereby an intrinsically nonperturbative quantity.
Alternatively, one may rewrite $\delta{Z_e}$ in $\alpha(0)$ scheme as
\bqa
\delta{Z_e}\big|_{\alpha(0)} &=& {1\over 2} \Delta{\alpha_{\rm had}^{(5)}}(M_Z)
+{1\over 2}{\rm Re}\, \Pi^{{AA}(5)}(M_Z^2)
\nn\\
&+&  {1\over 2}\Pi^{AA}_{\rm rem}(0)-{s_W\over c_W}
{\Sigma^{A Z}_T(0)\over M_Z^2},
\label{dZe:alpha(0):scheme}
\eqa
where $\Pi^{{AA}(5)}(M_Z^2)$ is the photon vacuum polarization from five massless quarks at momentum transfer
$M_Z^2$.  $\Delta{\alpha_{\rm had}^{(5)}}(M_Z)$, as a nonperturbative parameter that encapsulate
the hadronic contribution to photon vacuum polarization, can be determined from the measured $R$ values in
low-energy $e^+e^-$ experiments.
$\Pi^{AA}_{\rm rem}(0)$ represents the vacuum polarization from the $W$ boson, charged leptons and top quark at zero momentum transfer.
Note all these terms except $\Delta{\alpha_{\rm had}^{(5)}}(M_Z)$ can be computed reliably in perturbation theory.

Two other popular variants of the on-shell renormalization scheme are the $\alpha(M_Z)$ and $G_\mu$ schemes.
The corresponding charge renormalization constant can be adapted from \eqref{eq:ze:orig} into
$\delta{Z_e}\big|_{\alpha(M_Z)}=\delta{Z_e}\big|_{\alpha(0)}- {1\over 2} \Delta\alpha(M_Z)$,
and $ \delta Z_e|_{G_\mu}=\delta Z_e|_{\alpha(0)}-{1\over 2}\Delta r $, respectively.
$\Delta\alpha(M_Z)=\Pi^{AA}_{f\neq t}(0)-{\rm Re}\,\Pi^{AA}_{f\neq t}(M_Z^2)$,
and the expression for the oblique parameter $\Delta r $ can be found in \cite{Denner:1991kt}.
For latter convenience, we split the $\Delta\alpha(M_Z)$ into two different sources of contributions:
\bqa
\Delta \alpha\left(M_Z\right)=\Delta \alpha_q\left(M_Z\right)+\Delta \alpha_{\ell}\left(M_Z\right),
\label{Delta:alpha:split:into:two:terms}
\eqa
where $\Delta\alpha_q$ and $\Delta\alpha_{\ell}$ represent the contributions from the five light quarks and three charged leptons, respectively.
They are defined by
\begin{subequations}
\bqa
\Delta\alpha_{q}(M_Z)&=&\Pi^{AA}_{f=u,d,s,c,b}(0)-{\rm Re}\,\Pi^{AA}_{f=u,d,s,c,b}(M_Z),\phantom{xxx}\\
\Delta\alpha_{\ell}(M_Z)&=&\Pi^{AA}_{f=e,\mu,\tau}(0)-{\rm Re}\,\Pi^{AA}_{f=e,\mu,\tau}(M_Z).
\eqa
\end{subequations}

The QED coupling constants in the $\alpha(M_Z)$ and $G_\mu$ schemes become
\begin{subequations}
\bqa
\label{dZe:alpha(MZ):scheme}
&& \alpha\left(M_Z\right)=\frac{\alpha(0)}{1-\Delta \alpha\left(M_Z\right)},
\\
\label{dZe:Gmu:scheme}
&& \alpha_{G_\mu}=\frac{\sqrt{2}}{\pi}G_{\mu} M_W^2\left(1-\frac{M_W^2}{M_Z^2}\right).
\eqa
\end{subequations}
In contrast to the $\alpha(0)$ scheme, these two sub-schemes effectively resum some universal large (non-)logarithmic terms
arising from the light fermion loop and top quark loop.

\vspace{0.2cm}

\noindent{\color{blue}\it Numerical predictions.} To make concrete predictions,
we adopt the
following values for the masses of various particles~\cite{Workman:2022ynf}:
$M_H=125.25\:{\rm GeV}$, $M_Z=91.1876\: {\rm GeV}$, $M_W=80.377\: {\rm GeV}$, $m_t=172.69 \:{\rm GeV}$,
$m_e=0.5109989 \: {\rm MeV}$, $m_{\mu}=105.65837 \: {\rm MeV}$,
$m_{\tau}=1.77686 \: {\rm GeV}$.
The masses of the charged leptons are retained only when computing $Z_e$.
In all other situations, we keep the top quark massive and treat the remaining five quarks and all charged leptons as massless.

We adopt the following values for various couplings:  $\alpha(0)=1/137.035999$,
$G_{\mu}=1.1663787\times 10^{-5}~{\rm GeV}^{-2}$, and $\Delta\alpha_{\rm had}^{(5)}(M_Z)=0.02764$~\cite{Workman:2022ynf}.
The running strong coupling constant at the scale of the Higgs mass  is set to $\alpha_s(M_H)=0.115$~\cite{Bonciani:2015eua}.
With the aid of (\ref{dZe:Gmu:scheme}), we derive $\alpha_{G_\mu}=1/132.168$.
To calculate $\alpha(M_Z)$, we must know the values of $\Delta \alpha_q(M_Z)$ and $\Delta \alpha_l(M_Z)$.
As mentioned before, the nonperturbative parameter $\Delta \alpha_q(M_Z)$ can be interchangeably used with the experimentally
determined $\Delta\alpha_{\rm had}^{(5)}(M_Z)$.
$\Delta \alpha_{\ell}(M_Z)$ can be calculated in perturbation theory, whose value is determined through four-loop accuracy~\cite{Kuhn:1998ze,Eidelman:1995ny,Steinhauser:1998rq,Sturm:2013uka}.
We adopt the four-loop result $\Delta \alpha_{\ell}\left(M_Z\right)\approx 0.0314979$~\cite{Sturm:2013uka}.
Consequently, by employing (\ref{dZe:alpha(MZ):scheme}), we obtain $\alpha(M_Z)=1/128.932$.

At present the measured full width of the Higgs boson is subject to large uncertainty, $\Gamma_H=3.2^{+2.4}_{-1.7}$ MeV~\cite{Workman:2022ynf}.
When predicting the branching fraction of $H\to Z^0\gamma$,
we choose to use much more precise, theoretically predicted value of Higgs full width,
$\Gamma_H=4.07^{+4.0\%}_{-3.9\%}$ MeV~\cite{Workman:2022ynf,LHCHiggsCrossSectionWorkingGroup:2016ypw}, in order to reduce the theoretical error.

\begin{widetext}
\begin{table*}[htbp]\small
\centering
\setlength{\tabcolsep}{10pt}
	\renewcommand{\arraystretch}{1.6}
	{
		\begin{tabular}{ccccccc}
			\hline
		&
  $\alpha$ assignment at LO
  & $\Gamma^{\rm LO}$ & $\Gamma^{\mathcal{O}(\alpha)}$  & $\Gamma^{\mathcal{O}(\alpha_s)}$ & $\Gamma^{\rm Sum}$ & $ {\cal B}(\times 10^{-3})$  \\
			\hline
		$\alpha(0)$ scheme  & $\alpha^3(0)$	& $5.921$  & $0.313$   & $0.019$  & $6.252$  & $1.54\pm 0.06$   \\
			\hline
	$\alpha(M_Z)$ scheme & $\alpha(0)\alpha^2(M_Z)$	& $6.689$  & $-0.464$   & $0.021$  & $6.245$  & $1.53\pm 0.06$   \\
			\hline
	$G_\mu$ scheme 	& $\alpha(0)\alpha^2_{G_\mu}$ & $6.365$  & $-0.048$   & $0.020$  & $6.337$  & $1.56\pm 0.06$   \\
			\hline
	    Democratic scheme 	& $\alpha(0)\alpha(M_Z)\alpha_{G_\mu}$ & $6.525$  & $-0.249$   & $0.021$  & $6.297$  & $1.55\pm 0.06$   \\
			\hline
		\end{tabular}
\caption{Predicted values of the partial width (in units of keV) for $H\to Z^0\gamma$ in various $\alpha$ schemes,
at various levels of
perturbative accuracy.
$\Gamma^{\rm LO}$ represents the LO prediction to the decay width.
The contributions from the NLO electroweak and QCD corrections are denoted by $\Gamma^{\mathcal{O}(\alpha)}$ and $\Gamma^{\mathcal{O}(\alpha_s)}$, respectively.
We have taken the strong coupling constant $\alpha_s(M_H)=0.115$.
$\Gamma^{\rm Sum}$  is obtained by summing the LO contribution together with $\mathcal{O}(\alpha)$ and $\mathcal{O}(\alpha_s)$ corrections, and the rightmost column
is the branching fraction obtained via dividing $\Gamma^{\rm Sum}$ by the predicted full Higgs width, $\Gamma_H=4.07^{+4.0\%}_{-3.9\%}$ MeV~\cite{Workman:2022ynf,LHCHiggsCrossSectionWorkingGroup:2016ypw}.
\label{table:decay-width}	
}}
\end{table*}
\end{widetext}

In Appendix~\ref{appendix:A} we present the numerical results of the form factor $\widetilde{T}_5$ at NLO accuracy in different $\alpha$ schemes.
Plugging into \eqref{Partial:width:from:T5:form:factor}, we are able to present the predictions for the partial width in various $\alpha$ schemes.
It may look natural to take the QED coupling constant at the vertex attached to the outgoing photon as $\alpha(0)$.
For the electromagnetic coupling constant that appears elsewhere, one has some freedom to choose different $\alpha$ scheme.
In Table~\ref{table:decay-width} we enumerate our predictions for $\Gamma(H\to Z^0\gamma)$ (in units of keV)
based on four different $\alpha$ schemes at various levels of perturbative accuracy.
For reader's convenience, in Table~\ref{table:decay-width} we also juxtapose the contribution
of the NLO QCD correction to the partial width, which has been available long while ago~\cite{Spira:1991tj,Gehrmann:2015dua,Bonciani:2015eua}.

As can be seen from Table~\ref{table:decay-width}, we reconfirm that the NLO QCD correction is positive and tiny,
which constitutes only $0.3\%$ of the LO prediction of the partial width~\cite{Bonciani:2015eua}.
We also notice that, the NLO electroweak corrections is positive in the $\alpha(0)$ scheme,
while negative in other three schemes. In addition, the ${\cal O}(\alpha)$ corrections appear to be  sizable
in magnitude 
in both the $\alpha(0)$ and 
$\alpha(M_Z)$ schemes,
 which may approach approximately $^{+5\%}_{-7\%}$  of the LO prediction.
In contrast, the $G_\mu$ scheme as well as the Democratic scheme yield a relatively modest ${\cal O}(\alpha)$ correction.
Note the LO predictions of the partial width from four different $\alpha$ scheme are scattered in a wide range, from $5.921$ keV to $6.689$ keV.
Including the error of the Higgs full width, we find that the LO SM prediction for the branching fraction
lies in the range $(1.40-1.71)\times 10^{-3}$.
Interestingly, after including the NLO electroweak correction, the scheme dependence becomes substantially reduced,
with the relative uncertainties among different schemes less than $2\%$.
Taking the predictions from various $\alpha$ schemes as an estimation of the theoretical uncertainty, we then obtain the most accurate
SM prediction to be ${\cal B}[H\to Z^0\gamma]=(1.55 \pm 0.06)\times 10^{-3}$.
Note the error in the branching fraction mainly stem from the uncertainty associated with the full width of the Higgs boson~\footnote{We can estimate
other potential sources of uncertainty. The contribution from the top quark loop constitutes approximately $-10\%$
of the LO decay width. Taking into account that the Yukawa coupling strength of $Hb\bar{b}$ is suppressed with respect to $Ht\bar{t}$
by a factor of $m_b/m_t$,  we estimate that retaining the bottom quark mass may introduce a relative error of several per mille.
Furthermore, uncertainties about the $t$, $W$ masses may also introduce an uncertainty of several per mille  in the partial width.
All in all, these sorts of uncertainties are of the same order of magnitude as the NLO QCD correction,
which are significantly smaller than the uncertainty stemming from four different $\alpha$ schemes.}.

It is evident that the state-of-the-art SM prediction for the branching fraction is significantly lower than the measured value,
${\cal B}_{\rm exp}[H\to Z^0\gamma]=(3.4\pm 1.1)\times 10^{-3}$~\cite{ATLAS:2023yqk}. It 
may be too early to claim
that some sort of the BSM physics must be invoked to resolve this discrepancy.
More accurate measurements from {\tt HL-LHC}, and the prospective Higgs factories such as {\tt CEPC} and  {\tt FCC-ee},     will
play a crucial role to ultimately clear the smoke.

\vspace{0.2cm}

\noindent{\color{blue}\it Summary.} In this work we have calculated the NLO electroweak correction to the
rare decay process $H\to Z^0\gamma$ within the on-shell renormalization scheme.
To assess the theoretical uncertainty, we present the predictions of the partial width and the branching fraction
using several different $\alpha$ schemes.
In contrast to the tiny NLO QCD correction, the magnitude of the NLO electroweak correction turns out to become
sizable,  reaching $7\%$ of the LO result in the $\alpha(M_Z)$ schemes.
After including the ${\cal O}(\alpha)$ correction, the predictions from various $\alpha$ schemes tend to converge to each other,
which indicates that incorporating the NLO electroweak correction has significantly stabilized the SM prediction.
Our most accurate prediction is then ${\cal B}[H\to Z^0\gamma]=(1.55\pm 0.06)\times 10^{-3}$,
with the uncertainty predominantly stemming from the error of the full width of Higgs.
The relative error from varying the $\alpha$ schemes is less than $2\%$.
Although this state-of-the-art SM prediction is significantly lower than the measured value,
it is still premature to claim that one has to invoke some BSM scenarios to resolve this discrepancy.
Likely the problem mainly lies on the experimental side. Improved measurements of this rare decay process at {\tt HL-LHC},
and independent measurements at the prospective Higgs factories such as {\tt CEPC} and {\tt FCC-ee},
will be crucial.

\begin{acknowledgments}
We thank Yingsheng Huang for suggesting us to consider this project.
We are also indebted to Wen Chen and Yingsheng Huang for discussions.
Feynman diagrams in this work are drawn with the aid of {\tt JaxoDraw}~\cite{Binosi:2008ig}.
The work of W.-L. S. is supported by the NNSFC under Grant No.~12375079, and the
Natural Science Foundation of ChongQing under Grant No. CSTB2023 NSCQ-MSX0132.
The work of F.~F. is supported by the NNSFC under Grant No. 12275353.
The work of Y.~J. is supported in part by the NNSFC under Grant No.~11925506.
\end{acknowledgments}

\vspace{0.2 cm}

{\it Note added.} While we were finalizing this work, a preprint by Chen, Chen, Qiao and Zhu has recently appeared~\cite{Chen:2024vyn}.
Our results in $\alpha({M_Z})$ and $G_{\mu}$ schemes are compatible with their mixed 1 and mixed 2 schemes,
provided that the same input parameters are used.  However, our result in $\alpha(0)$ scheme slightly differs from theirs,
probably due to the different treatment of the light quark contribution to $Z_e$.

\appendix

\section{The expressions of $\widetilde{T}_5$  from four different $\alpha$ schemes}
\label{appendix:A}

Throughout this work we always retain the QED coupling associated with the photon emission vertex to be
$\alpha(0)$.  As aforementioned, there still  exists some freedom in the on-shell renormalization scheme to handle the charge renormalization.
In this appendix, we present the expressions of the form factor $\widetilde{T}_5$ in four different $\alpha$ schemes,
which differ in choosing the values of the QED coupling constant in other vertices.
Plugging these equations into \eqref{Partial:width:from:T5:form:factor},
we then obtain the predicted partial widths associated with each scheme, as enumerated in Table~\ref{table:decay-width}.

\subsection{$\alpha(0)$ scheme}
In this scheme, all the QED couplings in three vertices of the LO diagrams in Fig.~\ref{Feynman:Diagram:LO}
are chosen to be $\alpha(0)$, {\it i.e.}, the fine structure constant in the Thomson limit.
After including both ${\cal O}(\alpha)$ and ${\cal O}(\alpha_s)$ corrections,
the form factor $\widetilde{T}_5$ reads
\bqa
\label{T5:alpha(0):scheme}
&&\widetilde{T}_5=1.546\,\alpha^{3/2}(0)\bigg[1+\Delta{\alpha_{\rm had}^{(5)}}(M_Z)-0.535 \;\frac{\alpha(0)}{\pi}\nn\\
&&\quad\quad\quad \quad\quad\quad
+0.043\;\frac{\alpha_s}{\pi} \bigg],
\eqa
which depends on both charged leptons and light quarks, with the effects of the latter encapsulated in $\Delta{\alpha_{\rm had}^{(5)}}(M_Z)$.
As can be clearly seen, the prefactor accompanying  $\alpha(0)$ is about two orders-of-magnitude
greater than that accompanying $\alpha_s$,
which explains why the electroweak correction is much more important than the QCD correction.

\subsection{$\alpha(M_Z)$ scheme}

In this scheme, we choose the QED coupling constants associated with two vertices other than the photon emission vertex
in Fig.~\ref{Feynman:Diagram:LO} to be $\alpha(M_Z)$.
With the aid of \eqref{Delta:alpha:split:into:two:terms} and \eqref{dZe:alpha(MZ):scheme},
making the following substitution in (\ref{T5:alpha(0):scheme}):
\bqa
\label{eq:transfer-a0-amz}
\alpha(0)&\to& \alpha(M_Z)\bigg[1 -\Delta{\alpha}(M_Z)\bigg]\nn\\
&\to&\alpha(M_Z)\bigg[1 -\Delta{\alpha_{\rm had}^{(5)}}(M_Z)-\Delta \alpha_{\ell}(M_Z)\bigg]
\nn\\
&=&\alpha(M_Z)\bigg[1 - 0.02764- 13.527\,\frac{\alpha(M_Z)}{\pi}\bigg],\phantom{xx}
\eqa
we then obtain
\bqa
\label{T5:alpha(mz):scheme}
&&\widetilde{T}_5=1.546\,\alpha^{1/2}(0)\alpha(M_Z)\bigg[1 - 14.062 \;\frac{\alpha(M_Z)}{\pi}
\nn\\
&&\quad\quad\quad \quad\quad\quad
+0.043\;\frac{\alpha_s}{\pi}\bigg].
\eqa

To be compatible with the level of accuracy in the computation of  $\delta{Z_e}\big|_{\alpha(0)}$ in $\alpha(0)$ scheme,
we have utilized the one-loop expression of $\Delta \alpha_{\ell}(M_Z)$ in (\ref{eq:transfer-a0-amz}), which reads
\bqa
\Delta \alpha_{\ell}(M_Z)=\frac{\alpha(0)}{3\pi}\bigg(\ln\frac{M_Z^2}{m_e^2}+\ln\frac{M_Z^2}{m_\mu^2}+\ln\frac{M_Z^2}{m_\tau^2}-5\bigg).
\eqa
It might be worth emphasizing that $\widetilde{T}_5$ in (\ref{T5:alpha(mz):scheme}) is insensitive to the masses of all light quarks and charged leptons.

\subsection{$G_\mu$ scheme}

In this scheme, we choose the QED coupling constants affiliated with two vertices other than the
photon emission vertex in Fig.~\ref{Feynman:Diagram:LO} to be $\alpha_{G_\mu}$.
Making the following substitution in (\ref{T5:alpha(0):scheme}):
\bqa
\label{eq:transfer-a0-aGmu}
\alpha(0)&\to &
\alpha_{G_\mu}\bigg[1 -\Delta{\alpha_{\rm had}^{(5)}}(M_Z)+\Delta \alpha_{q}(M_Z)-\Delta r\bigg]
\nn\\
&=&\alpha_{G_\mu}\bigg[1 -0.02764- 1.030\,\frac{\alpha_{G_\mu}}{\pi}\bigg],\phantom{xx}
\eqa
we then obtain
\bqa
\label{T5:alpha(Gmu):scheme}
&&\widetilde{T}_5=1.546\,\alpha^{1/2}(0)\alpha_{G_\mu}\bigg[1- 1.565\;\frac{\alpha_{G_\mu}}{\pi}
\nn\\
&&\quad\quad\quad \quad\quad\quad
+0.043\;\frac{\alpha_s}{\pi}\bigg].
\eqa
Note that in \eqref{eq:transfer-a0-aGmu} we have utilized the one-loop values for $\Delta \alpha_{q}(M_Z)$ and $\Delta r$.
The explicit expression of $\Delta r$ is too complicated to present here, which can instead be found in
Eq.~(8.14) of \cite{Denner:1991kt}.
It is worth emphasizing that since both $\Delta \alpha_{q}(M_Z)$ and $\Delta r$ possess identical sensitivity to the light quark masses,
their difference in \eqref{eq:transfer-a0-aGmu} is not sensitive to nonperturbative hadronic effect, thus amenable to a perturbative treatment.
As in $\alpha(M_Z)$ scheme, $\widetilde{T}_5$ in $G_\mu$ scheme is also insensitive to the masses of light quarks and charged fermions.

\subsection{Democratic scheme}

In the Democratic scheme, we choose the QED couplings affiliated with three vertices in Fig.~\ref{Feynman:Diagram:LO} to be
$\alpha_{G_\mu}$, $\alpha(M_Z)$, and $\alpha(0)$, respectively.
By plugging (\ref{eq:transfer-a0-amz}) and (\ref{eq:transfer-a0-aGmu}) into (\ref{T5:alpha(0):scheme}),
we obtain
\bqa
\label{T5:mixed:scheme}
&&\widetilde{T}_5=1.546\,\alpha^{1/2}(0)\alpha^{1/2}(M_Z)\alpha_{G_\mu}^{1/2}\bigg[1- 7.814\times
\nn\\
&&\quad\quad\quad \quad\quad\quad
\;\frac{\alpha^{1/2}(M_Z)\alpha_{G_\mu}^{1/2}}{\pi}
+0.043\;\frac{\alpha_s}{\pi}\bigg].
\eqa


\end{document}